\documentclass[twocolumn,showpacs,preprintnumbers,amsmath,amssymb]{revtex4}

\usepackage{dcolumn}
\usepackage{bm}

\usepackage{amsmath}
\usepackage{amsfonts}   
\usepackage{graphics}  
\usepackage{psfrag} 
\usepackage{graphicx} 
\usepackage{epsfig}    
\usepackage{rotating}  
 \usepackage[latin1]{inputenc}
 \usepackage{color}
 \usepackage{rotating}

\usepackage{color}
\usepackage{epsfig}
\definecolor{darkgreen}{rgb}{0,0.5,0} 
\definecolor{violet}{rgb}{0.5,0,0.5}
\definecolor{orange}{rgb}{0.2,0.5,0.5}

\begin{document}

\preprint{}

\title{Evolutionary and Population Dynamics: A Coupled Approach 
}

\author{Jonas Cremer, Anna Melbinger, and Erwin Frey}
\affiliation{Arnold Sommerfeld Center for Theoretical Physics (ASC) and Center for NanoScience (CeNS), Department of Physics, Ludwig-Maximilians-Universit\"at M\"unchen, Theresienstrasse 37, D-80333 M\"unchen, Germany}

\date{\today}
\begin{abstract}

We study the interplay of population growth and evolutionary dynamics using a stochastic model based on birth and death events. In contrast to the common assumption of an independent population size, evolution can be strongly affected by population dynamics in general. Especially for fast reproducing microbes which are subject to selection, both types of dynamics are often closely intertwined. We illustrate this by considering different growth scenarios. Depending on whether microbes die or stop to reproduce (dormancy), qualitatively different behaviors emerge. For cooperating bacteria, a permanent increase of costly cooperation can occur. Even if not permanent, cooperation can still increase transiently due to demographic fluctuations. We validate our analysis via stochastic simulations and analytic calculations. In particular, we derive a condition for an increase in the level of cooperation. 
\end{abstract}
\maketitle

\section{Introduction}

The time evolution of size and internal composition of a population are both driven by discrete birth and death events. As a consequence, population dynamics and internal evolutionary dynamics are intricately linked. The biological significance of this coupling has previously been emphasized~\cite{Charlesworth:1971, Roughgarden, Roughgarden:1979, Ginzburg:1983,Mueller:1988, Charlesworth:1994, Cressman,Hauert3,Hauert:2008}. Those studies mostly employ density-dependent fitness functions to phenomenologically derive sets of coupled deterministic equations for the size and composition of populations in various ecological contexts. While those studies correctly describe the evolutionary dynamics of large population sizes, they do not account for stochastic effects arising at low population sizes. These demographic fluctuations are naturally described in the theoretical framework of stochastic processes based on elementary birth and death events as recently introduced~\cite{Melbinger:2010}. In particular, this approach allows one to explore the role of fluctuations in populations with a time-varying population size.

To understand such interdependence of population and evolutionary dynamics, it is instructive to first review the decoupled and deterministic formulations of both. \emph{Evolutionary game theory} is a well defined framework to describe the temporal development of different interacting traits or strategies~\cite{Maynard,Hofbauer}. It has been established as a standard approach to describe evolutionary dynamics if the fitness is frequency-dependent, i.e.~if the fitness of a certain strategy depends on the abundance of other strategies within the population. Within the most basic setup, well-mixed populations are assumed and the evolution of strategies is solely determined by fitness advantages. The temporal development of the abundance $x_S$ of a trait S follows a \emph{replicator dynamics}~\cite{Maynard,Hofbauer,Nowakbook},
\begin{equation}
\partial_t x_S=\left(\phi_S-\bar\phi \right)x_S.
\label{eq:replicator}
\end{equation}
A trait's abundance increases if its fitness $\phi_S$ exceeds the average fitness $\bar \phi $ of the population. The frequency dependence, with $\phi_S$ a function of the abundances $\vec x$ of all strategies, provoke non-linearities in Eq.~(\ref{eq:replicator}). Starting from this standard approach, many specific examples and extensions thereof have been studied~\cite{Hofbauer,Szabo,Nowakbook}. This comprises, for example, the prisoner's dilemma, the snowdrift-game and other games in well-mixed populations~\cite{axelrod-1981-211,Maynard,Hofbauer,Nowakbook}. It further ranges from the role of spatial arrangements and network interactions ~\cite{Szathmary:1987,NowakSpatial,Nowak:1994,Szab'o:2002,Hauert:2004,Ohtsuki2006,Pacheco,Santos,Roca:2009,Roca:2009a}  via cyclic dominance~\cite{Reichenbach:2006,Reichenbach, Antal,Reichenbach:2008,Claussen:2008,Cremer:2008,Cressman,Alava2008,Andrae:2010}, structured populations~\cite{Traulsen:2006a,Tarnita:2009a}, modified update-rules~\cite{roca:158701,Altrock:2009}, multi-player games~\cite{Gokhale:2010} and evolutionary algorithms~\cite{Levin} to the influence of internal and external fluctuations \cite{Nowak:2004,Blythe:2007,Galla:2009,Traulsen:2005,Traulsen:2006,Cremer:2009}. While these models consider a wide range of evolutionary aspects, they mostly rely on one key assumption, a decoupled, constant population size.

In contrast, \emph{population dynamics} focuses on the time evolution of the population size and how it is determined by environmental impacts like limited resources or seasonal variations.  The dynamics is typically described by differential equations of the form~\cite{Murray, Hastings:1997,Kot:2001} 
\begin{equation}
\partial_t N=\mathcal F\left(N;t\right),
\label{eq:pop_dyn}
\end{equation}
where $\mathcal{F}\left(N;t\right)$ may explicitly depend on time~\cite{Murray}. The most prominent example is logistic growth~\cite{Verhulst}. While a small population grows exponentially, the growth rate decreases with increasing population size due to limitations of resources and the population size is bounded below a maximum carrying capacity.

Illustrative examples of dynamical changes in the population size comprise bacterial and other microbial populations~\cite{Monod:1949,Velicer:2003p377,Hall-Stoodley:2004}: A surplus in nutrients or other metabolism related factors, can lead to an immediate and strong growth of the population while resource limitations or antibiotics and other detrimental factors can imply a stop in growth or even an abrupt death of single individuals. Even for only slightly varying environmental conditions, a fixed population size is thus rather the exception than the rule. 

But microbes not only show rich population dynamics, they are also subject to diverse evolutionary forces~\cite{Lenski:1991,Elena:2003,West:2006,Buckling:2009,Hibbing2010}.  Microbes live in interacting collectives of different traits. Evolution is ubiquitous and strong forms of frequency-dependence can be observed. Public good scenarios where a metabolically costly biochemical product is shared among individuals are of particular interest from an evolutionary perspective, see e.g.~\cite{Velicer:2003p377,West:2006,Frey:2010,Xavier:2011,Damore:2011}. This includes, for example, nutrient uptake, like disaccharides in yeast~\cite{Greig:2004,Gore:2009,MacLean:2010}, collective fruiting body formation~\cite{Strassmann:2000,Velicer:2009}, or the active formation of biofilms~\cite{Rainey2003,Hall-Stoodley:2004,Nadell:2009,Hibbing2010}. An example regarding iron uptake is considered below in more detail~\cite{Diggle,Buckling:2007,Kummerli:2009}. Furthermore, synthetical microbial systems have been considered~\cite{Chuang:2009,Chuang:2010}.

Motivated by these recent studies of microbial systems, we here investigate the consequences of such an interdependence between evolutionary and population dynamics.  Employing a  previously introduced theoretical approach~\cite{Melbinger:2010}, we study the influence of different growth scenarios in combination with demographic fluctuations.

The outline of this article is the following. In Section~\ref{Sec:Model} we discuss the stochastic dynamics and its deterministic approximation. Furthermore, we consider the limits in which the model maps to  standard (deterministic and stochastic) formulations of evolutionary dynamics. In Section~\ref{Sec:PD} we consider the dilemma of cooperation in growing populations. Here, an increase of cooperation can be observed which is analyzed in detail. In particular, we discuss the outcomes for two different growth scenarios, i.e. a reproduction-dynamics which either is balanced by death events or simply arrests in the stationary case. Finally, we close with a short conclusion in Section~\ref{Sec:Conclusion}.

\section{Coupling of Evolutionary and Population Dynamics\label{Sec:Model}}

\subsection{Microscopic Model}

\begin{figure}
\centering
\includegraphics[width=0.8\columnwidth]{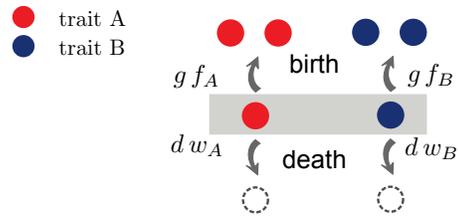}
\caption{(Color online) The per capita birth and death rates for two different traits, $A$ [light gray (red)] and $B$ [dark gray (blue)]. Each rate depends on a global, trait-independent and a relative, trait-dependent part. While the global and relative fitness terms, $g$ and $f_{A/B}$ affect the birth rates, the global and relative weakness terms $d$ and $w_{A/B}$ determine the death rates.  \label{Fig:Cartoon_rates}}
\end{figure}

We consider a population of $M$ different traits. Each trait $S$ is represented by $N_S$ individuals, such that the state of the population is given by $\vec N=(N_1,N_2,...,N_M)$. We further denote the frequencies of all different traits by $\vec x=\vec N/N$ with $N=\sum_S N_S$ being the total population size. The stochastic evolutionary dynamics is formulated in terms of per capita birth and death rates, $G_S$ and $D_S$, respectively.  The total rate for the abundance of trait $S$ to increase or decrease by one individual is given by
\begin{eqnarray}
\Gamma_{S \to 2S}=G_S N_{S},\hspace{0.5cm} \Gamma_{S\to\emptyset}&=&D_S  N_{S}.
\label{eq:rates}
\end{eqnarray}
The various biological factors determining each rate can be split up into two parts, a global and a relative contribution. While the global term is trait-independent and affects all traits in the same manner the relative term is trait-dependent and sets the differences between traits. We write
\begin{align}
G_S=g(\vec x,N)f_S(\vec x),\hspace{0.3cm}D_S=d(\vec x,N)w_S(\vec x)\label{eq:ratesdetail},
\end{align}
and refer to $g(\vec x,N)$ and $d(\vec x,N)$ as \emph{global birth-fitness} and \emph{global weakness}, respectively. The trait-dependent terms are the \emph{relative birth-fitness} $f_S(\vec x)$ and the \emph{relative weakness} $w_S(\vec x)$\footnote{In this work, we assume the relative parts to be independent of the system size. However, including a density dependent part also in the relative terms is straightforward.}. While birth-fitness terms affect the birth rates, weakness terms determine the expected survival times of individuals and hence their viability.  A short illustration of the stochastic processes is given in Fig.~\ref{Fig:Cartoon_rates} for the case of two different traits.

To specify the relative fitness terms, we follow the standard approach of evolutionary game theory~\cite{Maynard}, and assume them to depend linearly on the frequencies $\vec x$. Let  $\mathcal{P}$ be the payoff matrix for birth events. Then, the corresponding fitness vector for all traits is defined as
\begin{equation}
\vec \phi=1+s \mathcal{P} \vec  x \, .
\end{equation}
Following standard formulations, the \emph{selection strength} $s$ defines the relative weight of a frequency-dependent part with respect to a background-fitness set to 1~\cite{Nowak:2004,Traulsen:2005}. As will become clear in the following, it is convenient to make use of normalized fitness values, 
\begin{equation}
 \vec f=\vec \phi /\bar \phi,\label{eq:normfit}
 \end{equation}
where $\bar \phi = \sum_S \phi_S x_S$ such that $\bar f= \sum_S f_S x_S = 1$. Without loss of generality, this choice separates global and relative parts in such a way that the dynamics of the population size depends only on the global functions $g$ and $d$; see also the following Eq.~(\ref{eq:mfN}). An analogous approach with a payoff matrix $\mathcal{V}$ for death events can be used to obtain the frequency-dependent weakness functions $w_S$, which are also taken as normalized, $\sum_S w_S x_S=1$. 
Of course, a more general, non-linear frequency-dependance for both relative functions can readily be taken into account. For example, in microbial systems the fitness of an individual or the whole community depends in an intricate way on a plethora of factors, e.g.\ the abundance of individuals, secretion and detection of signaling molecules, toxin secretion leading to inter-strain competition and changes in environmental conditions. Non-linear frequency-dependent fitness-functions might help to account for such factors, see e.g.~\cite{Gore:2009,Frey:2010}.

In general, the global terms $g(\vec x,N)$ and $d(\vec x,N)$ depend on the population size and are frequency-dependent. Limited growth is one example of size-dependence. In such a setting, small populations start to grow exponentially but growth is bounded due to limited resources, e.g. $d(\vec x,N)$ increases with $N$. Frequency-dependent terms can, for example, occur in public good situations, as discussed in Section \ref{Sec:PD}.

\subsection{Dynamics}

The per-capita birth and death rates,~Eqs.~(\ref{eq:ratesdetail}), define a continuous-time Markov process~\cite{Gardiner,VanKampen:2001}. It is described by a master equation for the probability density $P(\vec N; t)$ to find the population in state $\vec N$ at time $t$:
\begin{align}
\frac{d P(\vec N; t )}{dt}=\sum_S  & \left[ (\mathbb{E}_S^--1)G_S N_S\, \right. \nonumber \\
& + \; \left.(\mathbb{E}_S^+-1)D_SN_S\right]P(\vec N; t ).
\label{eq:master1}
\end{align}
Here, $\mathbb{E}_S^\pm$ are step operators increasing/decreasing the number of individuals of trait $S$ by one~\cite{VanKampen:2001}, e.g.
\begin{eqnarray}
\mathbb{E}_S^\pm P(\vec N; t)=P\left(N_1,...,N_S\!\pm\! 1,...,N_M; t\right).\nonumber
\end{eqnarray}

For a reference it is instructive to first consider a deterministic limit where both fluctuations and correlations can be neglected. Then, upon factorizing higher moments of the probability density~\cite{Gardiner,VanKampen:2001}, one finds a closed set of equations for the expected frequencies $x_S$ and the total population size $N$:
\begin{subequations}
\begin{align}
\partial_t N =&\; \large[g\left(\vec x,N\right)\bar f-d\left(\vec x,N\right)\bar w\large]N\label{eq:mfN}\,, \\
\partial_t x_S =& ~~~~g(\vec x,N)\left[f_S(\vec x)-\bar f\,\right] x_S \nonumber \\
                       & -\, d(\vec x,N)\left[w_S(\vec x)-\bar w \right] x_S \label{eq:mfx}\,,
\end{align}
\label{meanfield}
\end{subequations}
where $\bar f=\bar w = 1$ according to Eq.~\eqref{eq:normfit}. To unclutter notation, we have not explicitly marked the expectation values in Eqs.~(\ref{meanfield}) but use the same notation as for the stochastic variables. 

This set of coupled non-linear equations resembles other deterministic approaches~\cite{Charlesworth:1971,Roughgarden,Roughgarden:1979,Ginzburg:1983,Charlesworth:1994,Hauert3,Cressman} and has a simple interpretation. Eq.~\eqref{eq:mfN} describes the population dynamics. As is typical for a deterministic approach, the dynamics does not depend on the global birth-fitness, $g$, and the global weakness, $d$, separately, but only on their difference. Eq.~\eqref{eq:mfx} describes the internal evolution of the population: The time evolution of the frequency of a strategy $S$ is given by the interplay between a growth and a death term. Each of them consists of a relative term measuring the surplus of the fitness/weakness relative to the corresponding population average. The weight of these terms are given by the respective global fitness functions, $g$ and $d$. During phases of population growth, where $g > d$ holds (see Eq.\eqref{eq:mfN}), the growth term and hence differences in relative birth fitness dominate the internal evolution of the population. Similarly, weakness differences are the main evolutionary driving forces during population decline.

From these considerations it follows that both the time scale of population and evolutionary dynamics have a crucial impact on the dynamics. This is obvious if the time-scales are similar. Such biological situations have been observed in many examples, see e.g.~\cite{Yoshida:2003,Hairston:2005,Saccheri:2006,Carroll:2007}. But also if evolution happens on longer time-scales than ecology this coupling can affect the evolutionary outcome as we show in the following.

Importantly, fluctuation cannot be ignored in general but can change evolutionary dynamics dramatically. Then, the deterministic approach given by Eqs.~(\ref{meanfield}) is not adequate. This regards for example fixation and extinction events but also the evolution of first and higher moments of a trait's abundance. For a proper description, one has to take the full stochastic dynamics and master equation~(\ref{eq:master1}) into account. One example, where fluctuations drastically change the outcome is given in the following Section~\ref{Sec:PD}.

\subsection{Mapping to Standard Approaches: Replicator Dynamics and the Moran Process\label{Sec:mapping}}

We now consider in which limits and to what extent our stochastic approach resembles the standard approaches of evolutionary dynamics. Let us first consider the special case where the global rates  $g(\vec x,N)\equiv g(N)$ and $d(\vec x,N)\equiv d(N)$ are frequency-independent and the ensuing deterministic dynamics exhibits a stable fixed point $N^*$ in the population size.  Then, birth and death events exactly balance each other, $g(N^*)=d(N^*)$, such that $N^*$ is fixed, $\partial_t N^* =0$. This is, for example, the case if the population size evolves according to a logistic growth law and the carrying capacity has been reached. In the deterministic limit, the internal dynamics, Eq.~\eqref{meanfield}(b), simplifies to
\begin{align}
\partial_t x_S &= g(N^*)\left[f_S(\vec x)-\bar f - w_S(\vec x)+\bar w \right]x_S.\label{eq:mfx_nstar}
\end{align}
\label{meanfield_simple}
The fraction $x_S$ evolves like in a standard replicator equation,  similar to~Eq.(\ref{eq:replicator}). It is the difference of both relative terms, the effective fitness $f_S-w_S$, which determines internal evolution. Compared to Eq.~\eqref{eq:replicator}, the additional constant prefactor $g(N^*)$ in Eq.~(\ref{eq:mfx_nstar}) just rescales the time-scale on which internal evolution occurs~\cite{Blythe:2007}. 

Furthermore, also the full stochastic formulations of our model and the standard stochastic approaches with a fixed population size resemble each other.  In those standard approaches, the birth of one individual is directly coupled to the death of another one. The dynamics is described by \emph{update rules}. For example, for the time-continuous formulation used here, the stochastic dynamics can be described by the Moran process~\cite{Moran,Ewens,Nowak:2004,Traulsen:2005,Blythe:2007,Traulsen:2006b,Traulsen:2006}~\footnote{Similarly, the stochastic dynamics is described by a Fisher-Wright process for discrete time-steps. Other update-rules are based on other fitness-functions or the way one individual replaces another one}. In our formulation, this process holds in the limit where the fixed point of the population size, $N^*$, is linearly stable with a large stability coefficient~\footnote{To strictly ensure $N$ to vary around $N^*$ with $\pm 1$, the fixed point has to be linear stable with additional higher orders supporting the stability.}. Then, a birth event is directly followed by a death event and vice versa. The effective rate for such a combined birth-death event is given by,
\begin{align}
\tilde\Gamma_{S \to S'}=\Gamma_{S' \to 2S'}\Gamma_{S\to \emptyset}+\Gamma_{S\to \emptyset}\Gamma_{S' \to 2S'}.
\label{eq:combined}
\end{align}
The strength of fluctuations in the fraction of a certain species is of the order $1/\sqrt{N^*}$ and the transition rate $\tilde\Gamma_{S \to S'}$ follows by the logic of an urn-model where, fitness-dependent, individuals reproduce to substitute other, randomly chosen, individuals~\cite{Moran,Ewens,Nowak:2004,Traulsen:2005,Blythe:2007}.

Beyond the Moran process, however, if $N^*$ is not linearly stable with sufficiently high stability coefficients, then birth and death events do not strictly follow each other. Depending on the stability of the fixed point, evolutionary paths deviating from $N^*$ by more than one individual have to be taken into account to derive an effective rate for a combined birth-death event.

In general, the population size changes with time, $N=N(t)$. For frequency-independent global rates, the deterministic limit of the internal evolutionary dynamics resembles the form of a replicator equation,
\begin{subequations}
\begin{align}
\partial_t N =& \, \large[g\left(N\right)-d\left(N\right)\large]N\label{eq:mfN_nofreq},\\
\partial_t x_S =&\, \lbrace g(N)\left[f_S(\vec x)-\bar f\,\right] 
\\&\, - d(N)\left[w_S(\vec x)-\bar w \right]\rbrace x_S.\label{eq:mfx_nofreq}
\end{align}
\label{meanfield_nofreq}
\end{subequations}
However, in contrast to Eq.~\eqref{eq:replicator}, both relative fitness terms, $f$ and $w$, are now weighted by the global rates. This has important implications. While in growth phases with $g>d$ the relative birth fitness $f_S$ dominates the dynamics, the relative weakness functions $w_S$ dominate during population-decline, $g<d$. Moreover, the time-varying population size also leads to a changing strength of fluctuations $\sim1/\sqrt{N(t)}$. In particular, when fitness differences are weak and the dynamics is close to neutral evolution, such a change might have strong consequences~\cite{Kimura,Traulsen:2005,Blythe:2007,Cremer:2009,traulsen_extime}.

\section{The Dilemma of Cooperation in Growing Populations}\label{Sec:PD}

To exemplify the importance of coupling and fluctuations offered by our approach, we here study the dilemma of cooperation in growing populations. This is motivated by the dynamics observed in microbial  biofilms where strong forms of cooperation can be observed~\cite{Velicer:2003p377,West:2006,Nadell:2009,Hibbing2010,Xavier:2011,Damore:2011}. Single individuals produce metabolically costly products which they release into the environment to support, for example, biofilm formation or nutrient depletion. As these products are available for other bacteria in the colony, the cooperating  individuals are producers of a public good, and, by having the extra load of production, permanently run the risk to be undermined by non-producing free-riding strains. An example is provided by the proteobacterium \emph{Pseudomonas aeruginosa}~\cite{Diggle,Buckling:2007,Kummerli:2009}. To facilitate the metabolically important iron-uptake, these microbes produce siderophores which they release into the environment. Given the high binding affinity to iron, these proteins are capable of scavenging single iron atoms from larger iron clusters. The iron-siderophore complex can then be taken up by the bacteria, ensuring their iron supply. However, as every bacterium, not only the producing ones, can take advantage of the released siderophores there is a dilemma of cooperation: While it would be optimal for the whole population to cooperate, cooperators are endangered due to their reproduction disadvantage.

In addition to the evolutionary dynamics, microbial colonies are also subject to strong changes in population size~\cite{Monod:1949,Stoodley:2002,Velicer:2003p377,Hall-Stoodley:2004}. While in the presence of nutrients, small colonies grow exponentially, growth is bounded due to limitations in resources or deteriorating environmental conditions. This includes insufficient amounts of nutrients, a lack of oxygen or a poisoning by metabolites. Eventually the colony size remains constant or even declines again~\cite{Monod:1949}.  Given by the exact interplay of these detrimental and other environmental factors, and differing from species to species, growth dynamics varies between two scenarios\cite{Jones:2010,Lennon:2011}. First, bacteria can switch into a dormant state where individuals stay alive but regulate reproduction rates and metabolic activity towards zero (\emph{dormancy scenario}). Depending on environmental conditions dormancy can increase survival chances. For example, in the presence of antibiotics, this downgraded metabolism  can make bacteria less vulnerable leading to persistence~\cite{Lewis,Balaban:2004,Kussell:2005,Jong:2011}, or dormancy might hedge a population against strongly fluctuation environments~\cite{Caceres:2003,Lennon:2011,Jong:2011}. Second, environmental conditions can lead to death rates increasing with the population size $N$ while birth rates are only slightly affected~\cite{JeremyS.Webb08012003}. The population, therefore, reaches a state of dynamical maintained population size with the death rates exactly balancing the birth rates (\emph{scenario of balanced growth}). In many populations, a situation in-between both scenarios is observed. In pathogenes like \emph{P. aeruginosa}, the fraction of individuals transferring to the dormancy state varies between 20\% and 80\%~\cite{1999COLE}. In the following we consider both scenarios and their impact on internal evolution separately.
 
\subsection{The Balanced Growth Scenario}

Let us first study the balanced growth dynamics where, in the stationary state, birth and death events are both present, but exactly balance each other such that the population size is about constant. We consider a population which consists of two traits, cooperators ($C$) and free-riders ($F$). The total number of individuals in the population is given by $N=N_C+N_F$ and the fraction of cooperators by $x\equiv x_C=N_C/N$. The relative birth fitness, $f_S$ ($\phi_S$, if not normalized), accounts for the reproduction disadvantage of cooperating individuals. We study the well-know prisoner's dilemma~\cite{Maynard}~\footnote{More generally we could also study other types of interactions like the snowdrift game. However, as we want to show the importance of population dynamics for supporting cooperation we chose the worst case scenario for cooperation, the prisoner's dilemma.}:
\begin{align}
\phi_C=&1+s(\tilde b x-\tilde c),\nonumber\\
\phi_F=&1+s\tilde b x,\nonumber\\
\bar\phi =&1+s(\tilde b-\tilde c)x.
\label{eq:_relativef_coop}
\end{align} 

As introduced in Section~\ref{Sec:Model}, the frequency-dependent part is weighted with the strength of selection $s$. Individuals obtain a benefit $\tilde b$ from direct interaction with cooperators, while only cooperating individuals have to pay the cost $\tilde c$ for producing the public good. 
For the resulting normalized fitness functions, $f_S=\phi_S/\bar\phi$, the inequality $f_C<f_F$ always holds; within the same population, the reproduction rate of cooperators is always smaller than the one of free-riders.

In the following, we take the payoff parameters to be constant, $\tilde c= 1$ and $\tilde b=3$. Then, $s$ directly sets the time scale of the internal evolution. The relative weakness is assumed to be trait-independent and constant, $w_C=w_F=1$; free-riders and cooperators have equal survival chances.

Furthermore, because cooperators are the producers of a public good, the overall growth condition of a population improves with a higher level of cooperation. We here choose the global birth fitness to increase linearly with the level of cooperation,
\begin{align}
g(x)=1+px.\label{eq:unfrozenrates}
\end{align}
The parameter $p$ scales the positive impact of the presence of public good on the population. In the scenario of balanced growth, we consider death rates increasing with the population size. For specificity, we assume logistic growth~\cite{Verhulst} and set
\begin{align}
d(N)=N/K.
\end{align}
$K$ scales the maximal size a population can reach (carrying capacity) as discussed in detail below.

The master equation \eqref{eq:master1} describing the full stochastic dynamics then takes the form
\begin{align}
\frac{d P(N_C,N_F)}{dt}=&\left[(\mathbb{E}_C^-\!-\!1)gf_C N_C\!+\!((\mathbb{E}_F^-\!-\!1)gf_FN_F \right. \nonumber \\
&+\left.(\mathbb{E}_C^+\!-\!1)d\,N_C\!+\!(\mathbb{E}_F^+\!-\!1)d\,N_C\right]\times\nonumber \\
&~P(N_C,N_F).
\label{Eq:ME}
\end{align}

\begin{figure}
\centering
\includegraphics[width=0.8\columnwidth]{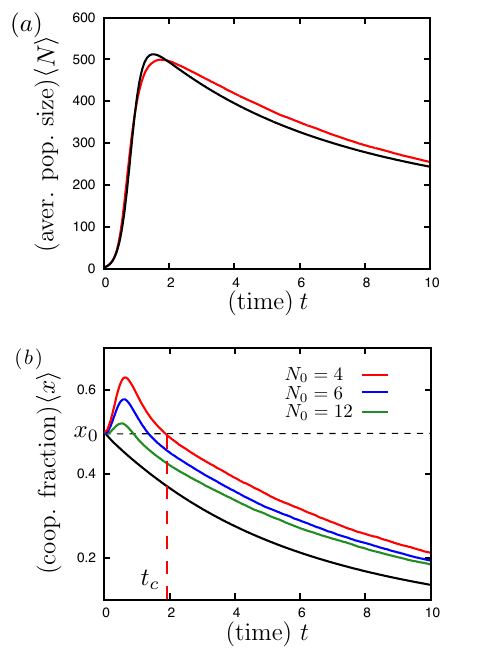}
\caption{(Color online) Cooperation in growing populations. Temporal development of ensemble averages. (a) The population size. Starting with $N_0=4$, the system grows exponentially until the carrying capacity is reached. It then falls again due to selection and a decreasing carrying capacity, see text. The full stochastic solution, gray (red) line, is described well by the deterministic approximation, black line. (b) The fraction of cooperators. It initially increases due to asymmetric amplification of fluctuations, and then falls again due to selection, see text. The level of cooperation, $x$, falls below its initial value, $x_0$ at the cooperation time, $t_C$. The transient increase is stronger for larger fluctuations and thus is stronger with a smaller initial population size $N_0$, see gray (colored) lines. The deterministic approximation do not account for this behavior, cf. black line. Parameters are $s=0.1$ and $p=10$.\label{Fig:unfrozen}}
\end{figure}

To explore the  dynamics, we performed extensive stochastic simulations. They were obtained by simulating $i=1, \ldots, R$ different realizations with the Gillespie algorithm~\cite{Gillespie}, according to the master equation \eqref{Eq:ME}. In Fig.~\ref{Fig:unfrozen}, we show the ensemble averages of the population size $\langle N \rangle$ and the fraction of cooperators $\langle x \rangle$ given by
\begin{subequations}
\begin{align}
\langle N\rangle&=\sum_i N_i(t)/R,\\
\langle x \rangle&=\sum_iN_{C,i}(t)/\sum_i N_i(t). \label{eq:ensemblex}
\end{align}
\label{eq:ensembleav}
\end{subequations}
This choice for the average naturally accounts for the fact that realizations with a larger populations size have a larger weight. It is especially important for biological situations where several realizations exist at the same time, e.g.~\cite{Chuang:2010}. In such an ensemble cooperation can increase in principle if there is a positive correlation between population size and the fraction of cooperators. The existence of this effect, also known as Simpson's paradox, has been shown recently by Chuang et al. for microbial populations~\cite{Chuang:2010}. Here we want to understand the dynamics underlying this correlation underlying cooperation. 

Starting with a small population, the system size grows exponentially (\emph{exponential phase}), reaches a maximum size and then declines again. Furthermore, and more strikingly, the disadvantage of cooperators can be overcome and a transient increase of cooperation can emerge. Even though the transient increase is caused by demographic fluctuations, it is instructive to examine the deterministic equations first. They not only describe the overshoot in the population size well, but also give insights into the relevant time scales of the dynamics:

\begin{subequations}
\begin{align}
\partial_t x&=-s(1+p x)x(1-x).\label{eq:pdx},\\
\partial_t N&=\left(1+px-\frac{N}{K}\right)N\label{eq:pdN}
\end{align}
\label{eq:pd}
\end{subequations}

The first equation describes the change in the average fraction of cooperators. The dynamics occurs on the time scale $\tau_x\sim1/s$, i.e. the strength of selection sets the time-scale of internal evolution. Note that $\partial_t x \leq0$ always holds and therefore the deterministic approximation cannot give rise to any transient increase of cooperation. In contrast, the dynamics of the total population size is well described deterministically, see~Fig.~\ref{Fig:unfrozen}(a). It resembles the well known equation of logistic growth~\cite{Verhulst} with a frequency-dependent maximal population size $K(1+px)$ (carrying capacity). During growth, changes in the population size occur on a time-scale $\tau_N\sim1+px$, c.f.~Eq.~\eqref{eq:pdN}. In the limit of weak selection, $\tau_N$ is comparably smaller than the time scale $\tau_x$, on which selection occurs. This and the frequency dependent carrying capacity are the reason for the overshoot: At the beginning the maximal population size is given by  $K(1+px_0)$. Because cooperators go extinct,  the size decreases with time. As this reduction is happening on a faster time scale than selection, $\tau_N<\tau_x$, the population size grows towards a larger carrying capacity, and then subsequently drops with decreasing carrying capacity due to a decline in cooperation.
\subsection{A Transient Increase of Cooperation}

The stochastic dynamics of the average fraction of cooperators $\langle x\rangle$ is qualitatively different from its deterministic limit. We observe a transient increase in the level of cooperation during a time window $(0,t_C)$. The magnitude of the \emph{cooperation time}, $t_C$,  strongly depends on the initial population size  $N_0$, cf. Fig.~\ref{Fig:unfrozen}. The origin of this transient increase in cooperation is a genuine stochastic effect: Demographic fluctuations during the initial phase are subsequently asymmetrically amplified by the population dynamics. Heuristically, this can be understood as follows; for a detailed mathematical analysis employing a van Kampen expansion see the next section.  

For a small initial population size $N_0$ demographic fluctuations are effectively symmetric, i.e., the occurrence of an additional cooperator or an additional free-rider are equally likely. However, the consequences of these two directions of demographic fluctuations differ strongly: In the exponential phase, an additional cooperator amplifies the growth of the population, while an additional free-rider hampers it.  In other words, fluctuations towards more cooperators imply a larger growth rate and hence a larger population size. Therefore, those realizations of the stochastic dynamics have a larger weight in the ensemble average, Eq.~\eqref{eq:ensemblex}, and enable an increase in the overall fraction of cooperators. With these considerations, a criterion for the transient increase of cooperation can be obtained: Demographic fluctuations, which are of size $\sqrt N$~\cite{Blythe:2007}, have to be large enough to overcome the selection pressure towards free-riders. This can already be inferred from Fig.~\ref{Fig:unfrozen}(b), where curves for three different values of the initial population size are shown. For the smallest $N_0$ the effect is the strongest because fluctuations are large at the beginning. In summary, a population bottleneck which corresponds to a small initial population size can favor cooperation transiently. Furthermore, if populations repeatedly undergo population bottlenecks, the increase in cooperation can be manifested also permanently.

\subsection{Van Kampen Expansion}

As discussed above the transient increase of cooperation is caused by fluctuations which are asymmetrically amplified. In order to quantify these findings analytically, we employ an Omega expansion in the system size according to van Kampen~\cite{VanKampen:2001} of the master equation~\eqref{Eq:ME}. For generality, we perform these calculations for arbitrary global growth function $g(x)$.
The deterministic solutions are separated from fluctuations by the following ansatz:
\begin{align}
N_C&=\Omega c(t) +\sqrt{\Omega}\xi, \nonumber\\
N_F&=\Omega f(t)+\sqrt{\Omega}\mu .\label {Eq:Substitution}
\end{align}
$c(t)$ and $f(t)$ correspond to the deterministic solutions, as shown below. $\xi$ and $\mu$ are fluctuations in the number of cooperators and free-riders. The relative strength of fluctuations and the deterministic parts are weighted by powers of $\Omega$ which scales with the current system size. For instance, to describe the transient increase which is generated by fluctuations at the beginning, $\Omega$ is given by $N_0$. Hence, this ansatz accounts for the fact that fluctuations scale as $1/\sqrt{N}$~\cite{Gardiner}.
Eq.~(\ref{Eq:ME}) is expanded in orders of $1/\sqrt{\Omega}$. With Eq.~(\ref{Eq:Substitution}), the step operators $\mathbb{E}_S^{+},~\mathbb{E}_S^{-}$ are given by
\begin{align}
\mathbb{E}_C^{\pm}=&1\pm \frac {1} {\sqrt{\Omega}} \partial_\xi+\frac{1}{2\Omega}\partial^2_\xi+\mathcal{O}(\Omega^{3/2}),\nonumber\\
\mathbb{E}_F^{\pm}=&1\pm \frac {1} {\sqrt{\Omega}} \partial_\mu+\frac{1}{2\Omega}\partial^2_\mu+\mathcal{O}(\Omega^{3/2}).
\end{align}
Employing these and Eqs.~(\ref{Eq:Substitution}) in Eq.~(\ref{Eq:ME}) leads to
\begin{align}
\partial_t P(\xi,\mu)-\sqrt\Omega(\dot c\partial_\xi+\dot f \partial_\mu)=
\left[-\sqrt\Omega g(\frac{\phi_C}{\bar\phi}c\partial_\xi+\frac{\phi_F}{\bar\phi}f\partial_\mu)\right.\nonumber\\
\left.+\Omega^0(...)+\frac{1}{\sqrt{\Omega}}(...)+\mathcal{O}(\Omega^{-3/2})\right]P(\xi,\mu),
\label{Eq:vanKampen}
\end{align}
where terms of the order $\Omega/K$ and higher are neglected. Initially, starting with a small population, these higher orders are very small because $\Omega\approx N_0$ and $N_0\ll K$ holds. The orders $\Omega^0$ and $1/\sqrt\Omega$ depend on $c,~f,~s,~\tilde b,~\tilde c,~ g,~\partial_\xi,~\partial_\mu,~\xi,~ \mu$ and are not written out in this equation for clarity. By collecting terms of order $\sqrt{\Omega}$ and using the identities $n=c(t)+f(t)$ and $x=c(t)/\left[c(t)+f(t)\right]$ the deterministic equations, Eqs.~(\ref{eq:pd}), are obtained (for $K\to \infty$). Higher orders of Eq.~(\ref{Eq:vanKampen}) lead to a Fokker-Planck equation for $P(\xi,\mu)$. From this Fokker-Planck equation, differential equations for the first and second moments of the fluctuations can be obtained. The first moments are given by 
\begin{align}
\langle \dot \xi \rangle= &\left[\frac{g\phi_C}{\bar\phi}+x(1-x)\partial_x\frac{g\phi_C}{\bar\phi}\right]\!\langle \xi \rangle-x^2\partial_x\frac{g(x)\phi_C}{\bar\phi}\langle \mu \rangle \nonumber\\
+&\!\frac{1}{2n\sqrt{\Omega}}\left[(1\!-\!x)^2\langle \xi^2\rangle\!-\!2x(1\!-\!x)\langle \xi \mu \rangle\!+\!x^2\langle \mu^2\rangle\right]\times\nonumber\\
& \partial_x^2\frac{g\phi_Cx}{\bar\phi},\nonumber \\
\langle \dot \mu \rangle=&(1-x)^2\partial_x\frac{g\phi_F}{\bar\phi}\langle \xi\rangle+\left[\frac{g\phi_F}{\bar\phi}-x(1-x)\partial_x\frac{g\phi_F}{\bar\phi}\right]\langle\mu\rangle \nonumber \\
+&\!\frac{1}{2n\sqrt{\Omega}}\left[(1\!-\!x)^2\langle \xi^2\rangle\!-\!2x(1\!-\!x)\langle \xi \mu \rangle\!+\!x^2\langle \mu^2\rangle\right]\times\nonumber\\
&\partial_x^2\frac{g\phi_F(1-x)}{\bar\phi}.\label{Eq:first_mom}
\end{align} 
Note that the second moments only couple at order $1/\sqrt{\Omega}$. Neglecting these higher orders, Eq.~(\ref{Eq:first_mom}) is linear and has an unstable fixed point at $(\xi,\mu)^*=(0,0)$.

Next, we analyze the impact of the second moments on the dynamics. Their coupling into Eq.~\eqref{Eq:first_mom} is only important for small times, when the first moments are still at the initial condition, the unstable fixed point $(\xi,\mu)^*=(0,0)$. Therefore, it is appropriate, to examine the second moments for small times,  $t\rightarrow0$. They then have the asymptotic form
\begin{align}
\partial_t \langle  \xi^2\rangle=&2ng\frac{\phi_C}{\bar\phi} x\nonumber,\\
\partial_t \langle  {\xi \mu} \rangle=& 0\nonumber, \\
\partial_t \langle  \mu^2 \rangle =&2ng\frac{\phi_F}{\bar\phi}(1-x).
\label{eq:2mom}
\end{align}
Due to the inhomogeneity of the differential equations, the second moments $\langle \xi^2 \rangle$ and $\langle \mu^2 \rangle$ immediatly start to grow. These non-zero second moments now couple back into the first moments, Eqs.~(\ref{Eq:first_mom}), and push them out of the unstable fixed point. To quantify this, the solution of Eqs.~(\ref{eq:2mom}) is employed in Eqs.~(\ref{Eq:first_mom}). The resulting equations are solved for small but finite times. As the increase of cooperation is caused by fluctuations, fluctuations have to establish first.  As fixed time we here consider the doubling time of the initial population $t_d=1/g(x)$. Within the time window $[0,t_d]$ evolution is neutral ($s\ll g(x)$)and thus $x=x_0$ holds. The approximation leads to a lower bound for the strength of fluctuations. Furthermore, the initial conditions are given by $\langle\xi_0\rangle=\langle \mu_0\rangle=\langle \xi_0^2\rangle=\langle \mu_0^2\rangle= 0$. If the initially generated and asymmetrically enhanced fluctuations are large enough to overcome the selection disadvantage, the transient increase of cooperation arises. To quantify this, the total fraction of cooperators in the system has to be examined:
 \begin{align}
 \frac{d}{dt}\langle x \rangle = \frac{\langle N_C\rangle }{\langle N_C+N_F\rangle}=& \frac{\dot x+1/\sqrt\Omega\langle \dot\xi\rangle}{n+1/\sqrt\Omega(\langle\xi\rangle+\langle\mu\rangle)}\nonumber\\
 &-\frac{(xn+1/\sqrt\Omega)(\langle\dot \xi\rangle+\langle\dot \mu)}{(n\sqrt\Omega+1/\Omega(\langle\xi\rangle+\langle\mu\rangle))^2}.
 \end{align}

For  $ \frac{d}{dt}\langle x \rangle>0$ the transient increase of cooperation is present. The condition $ \frac{d}{dt}\langle x\rangle=0$ leads,  to first order in $s$, to the transition line
\begin{equation}
s=\frac{\partial_x \ln[g(x)]}{n(1/g(x))\Omega}\Big |_{x_0}=\frac{\partial_x g(x)}{n(1/g(x))\Omega g(x)}\Big |_{x_0}.\label{eq:phase_transition}
\end{equation}

Here, $\Omega$ is given by the initial population size $N_0$.
For smaller $s$ there is a transient increase in cooperation, while for larger $s$ the level of cooperation decreases immediately. This resembles the condition for neutral evolution, e.g.~\cite{Kimura, Cremer:2009}; evolution is only neutral for $sN\lesssim const$. Thus, only if fluctuations are strong during the initial phase of the dynamics, such that the system behaves neutrally, they are sufficient to overcome the selection pressure towards free-riders.
The phase boundary and thereby the strength of the transient increase depends on $\partial_x g(x) |_{x_0}$ and $g(x_0)$. Both terms have antagonistic impacts on the transition line. The reason for this behavior is that the initial doubling time, i.e. the time during which fluctuations are the most pronounced,  decreases with increasing $g(x_0)$. The positive enhancement relies on the growth advantage of more cooperative realizations, which depends on $\partial_x g(x) |_{x_0}$ at the beginning. Note, that for non-linear growth functions, where $\partial_x g(x) |_{x_0}$ also depends on $x_0$, the transient increase can even be reduced by accounting for higher orders. This behavior was also experimentally observed in recent studies with microbes, where the growth advantage of cooperators was tuned~\cite{Chuang:2010}.  In the next paragraph, we show that the calculated phase boundaries match our simulation results very well for several distinct global growth functions.
\subsection{Phase Diagrams}

\begin{figure}
\centering
\includegraphics[width=0.8\columnwidth]{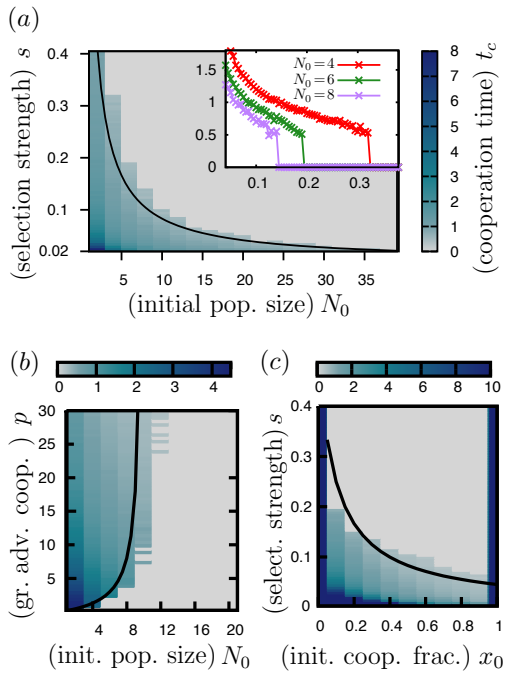}

\caption{
(Color online) The transient increase of cooperation and its dependence on parameters. Encoded in gray (colored) scale, the cooperation time $t_C$ is  plotted for three different pairs of parameters: $\lbrace N_0, s\rbrace$, $\lbrace N_0, p\rbrace$, and $\lbrace x_0, s\rbrace$ in (a), (b), and (c) respectively. The boundary between the regimes of transient increase and immediate decrease are in good agreement given by Eq.~(\ref{eq:phase_transition}), plotted as black lines. In the inset of (a), the cooperation time is shown for varying selection strength $s$: $t_C$ sharply drops at the boundary. Not varied parameters are given by $p=10,x_0=0.5$ in (a); $s=0.05$ and $x_0=0.5$ in (b); $p=10,N_0=6$ in (c).
\label{Fig:phase_unfrozen}}
\end{figure}

In the following we consider how the duration $t_C$ of the transient increase in cooperation depends on the system parameters  for the specific global growth function $g(x)=1+px$, cf. Fig.~\ref{Fig:phase_unfrozen}. Then, the transition line between a transient increase, $t_C>0$, and an immediate decrease, $t_C=0$, given by Eq.~\eqref{eq:phase_transition}, now reads,
\begin{equation}
s=\frac{p}{n\Omega(1+px_0)},\label{eq:phase_transition_linear}
\end{equation}
where $n\Omega=2N_0$.
For smaller selection strength, $s<\frac{p}{n\Omega(1+px_0)}$, the asymmetric amplification of fluctuations is sufficient to overcome the selection disadvantage of cooperators while for larger selection strength, $s>\frac{p}{n\Omega(1+px_0)}$, free-riders prevail.

In Fig.~\ref{Fig:phase_unfrozen}, we compare this result of the analytical calculations with the stochastic simulations. We observe that upon increasing the strength of selection, $s$, which sets the advantage of  free-riders, the cooperation time $t_C$ decreases. In contrast, stronger demographic fluctuations, their strength scales as $1/\sqrt{N_0}$, prolong the duration of the transient increase, i.e., $t_C$ increases with decreasing $N_0$, cf. Fig.~\ref{Fig:phase_unfrozen}(a). These two antagonistic effects lead to a sharp phase boundary between the regimes of transient increase ($t_C >0$) and immediate decrease ($t_C=0$); see inset of Fig.~\ref{Fig:phase_unfrozen}(a). Here, the cooperation time steeply drops to zero if  the strength of selection exceeds a critical value. The boundary line is in good agreement with Eq.~(\ref{eq:phase_transition_linear}), cf. black line in Fig.~\ref{Fig:phase_unfrozen}(a). 

In Fig.~\ref{Fig:phase_unfrozen}(b), the cooperation time is shown for varying initial population size $N_0$ and strength of the global fitness advantage due to cooperators, $p$. Now, the phase boundary is determined by the interplay between the size of demographic fluctuations and its amplification due to the global fitness advantage of more cooperative populations. $N_0$ has to be small enough for the asymmetric amplification mechanism to be effective. Again, the phase boundary is in good agreement with Eq.~(\ref{eq:phase_transition}); see solid black line in Fig.~\ref{Fig:phase_unfrozen}(b).

In Fig.~\ref{Fig:phase_unfrozen}(c), the cooperation time is plotted for varying initial cooperator fraction, $x_0$,  and selection strength, $s$. We find that the cooperation time decreases with increasing $x_0$. Remarkably, for small $x_0$, the amplification mechanism is especially pronounced and therefore able to compensate comparably large selection strengths $s$. This is again well described by Eq.~(\ref{eq:phase_transition_linear}), see Fig.~\ref{Fig:phase_unfrozen}(c)  (solid black line). The observation is of possible relevance for the evolution of cooperation since it allows one a small initial fraction of cooperators to proliferate in the population.

Taken together, our analytical calculations provide a mechanistic understanding for the transient increase of cooperation and its dependence on the system parameters $s$, $p$, $x_0$, and $N_0$. We have quantitatively calculated the phase boundary and gained insights into the basic nature of the transient increase: First, the probability distribution in the cooperator fraction $\langle x\rangle$ is broadened due to neutral evolution; note that Eq.~(\ref{eq:phase_transition_linear}) resembles the condition for neutral evolution~\cite{Kimura,Cremer:2009}. Second, these initially generated fluctuations are asymmetrically amplified and can, therefore, cause an increase in the level of cooperation.

\subsection{The Dormancy Scenario}

Let us now consider the dormancy scenario where the ability to reproduce decreases with increasing population size. For specificity, we assume the global birth and death functions to be given by
\begin{align}
g(x,N)=1+px-\frac{N}{K}, \hspace{.15cm}\text{and}~d=0.\label{eq:frozenrates}
\end{align}
In this scenario individuals do not die but the birth rates decrease towards zero as the population size reaches its carrying capacity. The relative functions, $f_S$ and $w_S$, are the same as before; the weakness terms are constant and the fitness terms given by Eq.~\eqref{eq:_relativef_coop}.

\begin{figure}
\centering
\includegraphics[width=0.8\columnwidth]{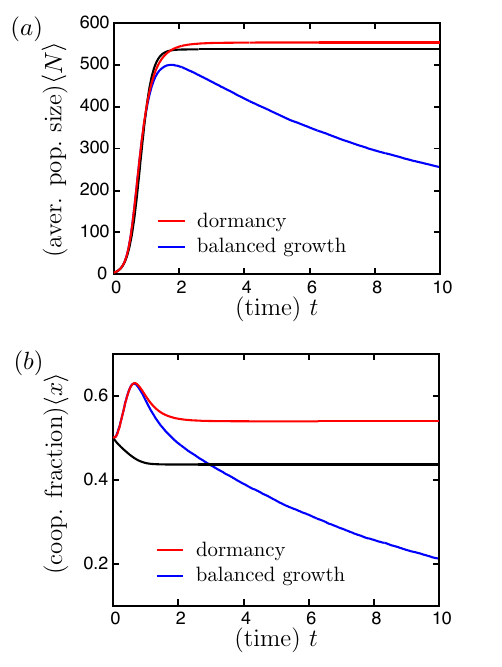}
\caption{(Color online) The dilemma of cooperation  in the dormancy scenario. (a) The growth dynamics. Initially, the small population grows exponentially until growth is stopped, cf. light gray (red) line. This behavior is well described by the deterministic equation \eqref{eq:pdN_frozen}, see black line. In contrast, for the balanced growth scenario, the dynamics continue and, due to selection, the population size falls again, see dark gray (blue) line. (b) The fraction of cooperators. Equal to the balanced growth scenario, dark gray (blue) line, there is an initial increase of cooperation due to asymmetric amplification within the dormancy scenario. Again, this is not described by the deterministic approximation, Eq.~\eqref{eq:pdx_frozen}. However, in contrast to the balanced growth scenario, the higher level of selection is latter fixed due to the stop in growth dynamics. Parameters are given by $s=0.05$, $p=10$, and $N_0=4$.
\label{Fig:frozen}}
\end{figure}

To understand the differences in the evolutionary outcome, we again study the deterministic rate equations first. They are given by
\begin{subequations}
\begin{align}
\partial_t N&=\left( 1+px-\frac{N}{K}\right)N\label{eq:pdN_frozen},\\
\partial_t x&=-s\left(1+p x-\frac{N}{K}\right)x(1-x)\label{eq:pdx_frozen}.
\end{align}
\label{eq:pd_frozen}
\end{subequations}
The equation describing population growth is formally identical to the corresponding equation in the balanced growth scenario, Eq.~\eqref{eq:pdN}. Differences arise because in the present case there is mutual feedback between internal and population dynamics. This coupling implies that both arrest once the population size reaches its carrying capacity. In the arrested state there is a relation between population size $N^*$ and composition $x^*$: $1+p x^*=N^*/K$. Thus, the reached stationary state, $(x^*,N^*)$, depends on the initial values $x_0$ and $N_0$. The precise mapping depends on the selection strength $s$. For weak selection (small $s$), the population dynamics is much faster than the internal dynamics and hence the population size reaches a stationary state while the composition is still at its initial value $x_0$, i.e., $N^* = K (1 + p x_0)$. In contrast, for strong selection, cooperators go extinct quickly with $x^*=0$ such that the stationary population size becomes $N^* = K$. An example for the deterministic dynamics is shown as a solid black line in Fig.~\ref{Fig:frozen}.  
As for balanced growth, the deterministic dynamics exhibits a strictly monotonous decrease in the cooperator fraction, with the difference that now the asymptotic value is arrested at some finite value. These differences are also reflected in the stochastic dynamics, where the asymmetric amplification mechanism is acting, cf. Fig.~\ref{Fig:frozen}. In the initial phase of the dynamics, this mechanism affects the time evolution of the cooperator fraction in the same way as for balanced growth, namely it leads to an initial increase of cooperation. Differences in birth and death rates, Eq.~\eqref{eq:unfrozenrates} and Eq.~\eqref{eq:frozenrates}, are negligible for small population size. The arrest of the dynamics only becomes effective at later times where an increase in population size implies a significantly declining birth rate. As a consequence even the stochastic dynamics becomes arrested such that the initial rise in the cooperator fraction may become manifested as a permanent increase. This will be the case if the dynamics becomes arrested during the time window where the asymmetric amplification mechanism acts; see red line in Fig.~\ref{Fig:frozen}(b).

\begin{figure}
\centering
\includegraphics[width=\columnwidth]{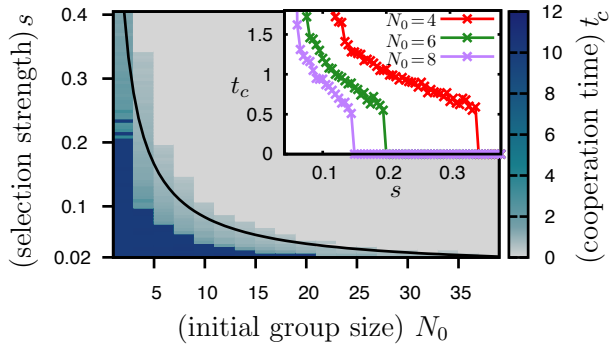}
\caption{
(Color online) The transient increase of cooperation for the dormancy scenario. The cooperation time $t_C$ depending on the initial population size, $N_0$, and the strength of selection, $s$. The condition for a transient increase of cooperation to occur is still given by Eq.~\eqref{eq:phase_transition}, black line. In addition, due to the stop in growth dynamics, there is an additional regime, where the increase becomes permanent, dark gray (dark blue) area. The permanent increase is also shown in the inset, where the cooperation time is shown for varying strength of selection. If, for a given initial population size, selection is sufficiently slow compared to fixation of the growth dynamics, the increase of cooperation becomes permanent. Parameters are given by  $p=10,~x_0=0.5$.
\label{Fig:phase_frozen}}
\end{figure}

In summary, there are now three scenarios for the dynamics, cf. Fig.~\ref{Fig:phase_frozen}. In addition to the immediate decline and transient increase there is now also a \emph{permanent increase} in the cooperator fraction. The analytical expression separating the regimes of transient increase and immediate decline still holds, Eq.~\eqref{eq:phase_transition_linear}, because it is due to the same mechanism as before. We did not manage to derive an explicit expression for the transition line to permanently increase. However, as the existence of a permanent increase in the cooperator fraction depends on the asymmetric amplification mechanism, the regime of permanent increase is bounded by a hyperbolic line beneath the one given by Eq.~\eqref{eq:phase_transition}. The latter is a necessary but not a sufficient condition for the permanent increase to occur.

\section{Conclusion}\label{Sec:Conclusion} 

In this article, we have given a synthesis of evolutionary and population dynamics. This is based on the understanding that birth and death events are the driving forces underlying changes in the size as well as the composition of a population~\cite{Melbinger:2010}. Both processes are inherently stochastic and inevitably lead to demographic fluctuations whose magnitude depends on the population size. The ensuing stochastic formulation thereby naturally accounts for the coupling between internal evolutionary dynamics and population dynamics. The evolutionary outcome of the dynamics is determined by the interplay between selection pressure and random drift caused by demographic fluctuations. Since our approach allows to study evolutionary dynamics with varying population size we can explore ecological situations where the relative impact of deterministic and stochastic evolutionary forces change with time. Thereby demographic fluctuations may lead to a dynamics which is qualitatively different from the corresponding deterministic dynamics: Beyond creating a broad distribution in size and composition, the coupling  can strongly distort the distribution and thus strongly influence average values. For the public good scenario, discussed in this paper, this corresponds to an asymmetric amplification mechanism which yields a transient increase in the level of cooperation.

In the absence of a  coupling between internal evolution and population dynamics, the impact of population size on the internal evolutionary dynamics reduces to a modulation in the strength of demographic fluctuations. If, in addition, the deterministic population dynamics exhibits a strongly attractive fixed point at a finite population size, our model maps to a  standard description of evolutionary dynamics, i.e. the Moran process.

The general observations made for the coupled stochastic dynamics are exemplified by the dilemma of cooperation in growing populations. Here, fluctuations in combination with growth lead to a transient increase of cooperation. The origin of this increase is the asymmetric amplification of fluctuations. As the presence of cooperators increases the growth rates, fluctuations towards those are enhanced. Therefore  growth dynamics cannot be ignored but can be an essential part in evolution. Furthermore, the details of the growth dynamics can be crucial in determining the evolutionary outcome. As we have considered for the dilemma of cooperation and two extremes of microbial growth dynamics, cooperation can either increase only transiently or the higher level can even fixate due to dormancy. Our analytical derived transition line provides the same sufficient condition for the transient increase in both scenarios.  Furthermore, the same line is also a necessary condition for the permanent increase for the dormancy scenario. In actual populations, both scenarios are present with a  fraction of 20\% - 80\% dormant bacteria~\cite{1999COLE}. While the transient increase does not depend on this fraction, the permanent increase  is smaller than for purely dormant bacteria. The discussed scenarios for the increase of cooperation,  rely on demographic fluctuations which are especially pronounced during population bottlenecks. Such bottlenecks  may be caused by seasonal changes of the environment, migration into new habitats and range expansion, e.g.~\cite{Hallatschek:2007,Nadell:2010,Hallatschek:2011, Kuhr:2011,Cremer:2011,Murray}. In addition, if the permanent increase is not present, repeated bottlenecks provoking regular occurring growth phases can favor cooperative behavior by stabilizing a former transient increase. This becomes especially important in the context of biofilms where population structure and involved restructuring mechanisms can drastically change evolutionary outcome~\cite{Rainey2003,Chuang:2009, Chuang:2010}.

\acknowledgements{We thank Jan-Timm Kuhr  for discussion. Financial support from the Deutsche Forschungsgemeinschaft through the SFB TR12 ``Symmetries and Universalities in Mesoscopic Systems" and the Nano Initiative Munich (NIM) is gratefully acknowledged.}

\bibliographystyle{apsrev}

\end{document}